\documentclass[12pt]{iopart}

\usepackage{graphicx}

\begin{document}

\title[Dressing a black hole with non-minimally coupled
scalar field hair]{Dressing a black hole with
non-minimally coupled scalar field hair}

\author{Elizabeth Winstanley}

\address{
Department of Applied Mathematics,
The University of Sheffield,
Hicks Building,
Hounsfield Road,
Sheffield.
S3 7RH
U.K.}

\ead{E.Winstanley@sheffield.ac.uk}

\begin{abstract}
We investigate the possibility of dressing a four-dimensional black hole with
classical scalar field hair which is non-minimally coupled to the space-time curvature.
Our model includes a cosmological constant but no self-interaction potential for the scalar
field.
We are able to rule out black hole hair except when the cosmological constant
is negative and the constant governing the coupling to the Ricci scalar curvature
is positive.
In this case, non-trivial hairy black hole solutions exist, at least some of
which are linearly stable.
However, when the coupling constant becomes too large, the black hole hair becomes
unstable.
\end{abstract}

\pacs{04.70.Bw, 04.20.Jb, 04.40-.b}



\section{Introduction}
\label{sec:intro}

The existence or non-existence of scalar field hair for black holes in
general relativity has been a subject of investigation for over thirty years
(see, for example, \cite{heuslerbook} for a detailed review, or \cite{bek96} for
a nice summary of the situation).
In this paper we are concerned with scalar field hair for black holes in
four-dimensional theories with the following action:
\begin{equation}
S=\int d^{4}x \, {\sqrt {-g}} \left[
\frac {1}{2}\left( R -2\Lambda \right)
-\frac {1}{2} \left( \nabla \phi \right) ^{2}
-\frac {1}{2} \xi R \phi ^{2} -V(\phi ) \right] ,
\label{eq:action}
\end{equation}
where $R$ is the Ricci scalar curvature, $\Lambda $ the
cosmological constant (which may be positive, negative or zero),
$\xi $ is the coupling constant, $V(\phi )$ the scalar field
self-interaction potential and $\left( \nabla \phi \right) ^{2} =
\nabla _{\mu } \phi \nabla ^{\mu } \phi $. For a minimally coupled
scalar field, $\xi =0$, and for conformal coupling, $\xi =1/6$ in
four dimensions.

The work to date in the literature has tended to concentrate on
the case of a minimally coupled scalar field in asymptotically
flat space-time, with a succession of no-hair theorems proved for
ever more general self-interaction potentials $V$
\cite{bek72a}-\cite{bek95}.
Although Bekenstein's original result
\cite{bek72a} (which ruled out hair for a massive scalar field
with no additional self-interaction potential) was for general
static space-times without the additional assumption of spherical
symmetry, subsequent work (such as
\cite{heusler95}-\cite{bek95}) assumed that the geometry was
spherically symmetric.
We shall also consider only static,
spherically symmetric space-times in this paper.

Dropping the assumption that the geometry is asymptotically flat, and introducing
a cosmological constant, the picture changes, with non-trivial, minimally coupled,
scalar field hair existing for certain non-zero potentials \cite{toriiAdS}-\cite{zlosh}.
However, if the self-interaction potential $V$ is identically zero, then
there is no scalar field hair independent of the sign of the cosmological constant
\cite{toriiAdS,toriidS}.
For non-zero self-interaction potential,
when the cosmological constant is negative,
at least some of the non-trivial hairy black
holes found are linearly stable \cite{toriiAdS}, but, when the cosmological constant
is positive, all solutions examined so far are unstable \cite{toriidS}.
Black holes with minimally coupled
scalar hair in asymptotically anti-de Sitter space have also
been of recent interest in supergravity \cite{hertog1},
but these solutions are unstable \cite{hertog2}.

Conformally coupled scalar field hair for black holes has been studied for as long
as minimally coupled scalar field hair, with the discovery of an exact, closed form,
solution of the field equations with a zero self-interaction potential,
known as the BBMB black hole \cite{bek74}-\cite{bbm}.
While this is the unique static, asymptotically flat, solution of the Einstein-conformal
scalar field system \cite{xanth}, the scalar field diverges on the event horizon,
and, furthermore, the solution is unstable \cite{bronnikov}.
The BBMB black hole solution can be
generalized in the presence of a cosmological constant, but only with a
non-vanishing self-interaction potential.
When the cosmological constant is positive, the corresponding solution has
a quartic potential \cite{martinez}, but is again unstable \cite{phil}.
For positive cosmological constant, there are no non-trivial
solutions when the self-interaction
potential is zero \cite{ewc}.
When the cosmological constant is negative, stable, non-trivial hairy black holes
have been found numerically when the potential is zero or quadratic \cite{ewc},
and there is also an exact, closed-form solution for a quartic potential
\cite{mtz}.

Non-minimal couplings other than conformal coupling have received less
attention in the literature, and, to date, only asymptotically flat
space-times have been considered.
No-hair theorems have been proved by various authors \cite{mayo}-\cite{saa2}.
Mayo and Bekenstein \cite{mayo} considered very general potentials, and
found it necessary to employ energy arguments to prove their no-hair
theorems.
The particular case of a quartic self-interaction potential has been considered
by Ay\'on-Beato \cite{ayon}, who proved a general no-hair theorem for
arbitrary coupling $\xi $.
For vanishing potential, there is strong numerical evidence \cite{pena}
that there are no non-trivial solutions for any coupling $\xi $ to the scalar curvature,
and the theorems of Saa \cite{saa1,saa2} back this up, although they make assumptions
about the scalar field which are necessary to employ the conformal transformation
technique \cite{bek74,maeda}, which we shall consider later in section
\ref{sec:conformal}.

The state of current knowledge of the existence and stability of scalar field
hair when the self-interaction potential is zero can therefore
be summarized in the following table:
\begin{center}
\begin{tabular}{|c||c|c|c|c|c|}
\hline
V=0 & $\xi <0 $ & $\xi =0$ & $0< \xi < 1/6$ & $\xi =1/6$ & $\xi > 1/6 $
\\
\hline
\hline
\hline
$\Lambda >0$ & & No hair & & No hair &
\\
\hline
$\Lambda =0$ & No hair & No hair & No hair & Unstable hair & No hair
\\
\hline
$\Lambda <0$ & & No hair & & Stable hair &
\\
\hline
\end{tabular}
\end{center}
Our purpose in this paper is to investigate the effect of introducing a
cosmological constant into these models, and we consider arbitrary coupling constant $\xi $
but a vanishing potential for simplicity.

The outline of this paper is as follows.
In section \ref{sec:model} we introduce our model, and
we also discuss the boundary conditions satisfied by the scalar field,
particularly at infinity, which will be crucial for our later analysis.
Next, in section \ref{sec:nohair}, we prove some simple no-hair theorems
in the cases $\Lambda >0$, $\xi >0 $ and $\Lambda <0 $, $\xi <0$,
using the technique of Bekenstein \cite{bek72a}.
In order to study the remaining cases, that is: $\Lambda >0$, $\xi <0$ and
$\Lambda <0$, $\xi >0$, in section \ref{sec:conformal}
we employ a conformal transformation \cite{bek74,maeda},
which maps the system with a non-minimally coupled
scalar field to one in which there is a scalar field which is minimally coupled
to the geometry but has a complex self-interaction.
This will enable us to prove a no-hair theorem in section \ref{sec:confapp}
for the case when $\Lambda >0$ and $\xi <0$.
For the remaining region of the $\Lambda, \xi $ parameter space,
when $\Lambda <0$ and $\xi >0$, we are unable to prove a no-hair theorem
and, in section \ref{sec:solns}, we find numerically
non-trivial hairy black holes in this case, before
examining the linear stability of these solutions in
section \ref{sec:stab}.
For the particular black hole solutions we study numerically, we
find that those solutions when $0<\xi <3/16$ are stable, whereas those
when $\xi >3/16$ turn out to be unstable.
Our conclusions are presented in section \ref{sec:conc}.

Throughout this paper, the metric has signature $(-+++)$,
and we will use units in which the gravitational coupling
constant $\kappa ^{2}=8\pi G$ (with $G$ being Newton's constant)
is set equal to unity and $c=1$.

\section{The Model}
\label{sec:model}

We consider the action (\ref{eq:action}), but
in this paper we shall consider only the simplest case, when the
self-interaction potential $V(\phi )\equiv 0$.
However, we shall comment in various places on whether our results for
$V\equiv 0$ are likely to be extendible to non-zero potentials.

Setting $V=0$ in (\ref{eq:action}) and then taking the variation
yields the Einstein field equations:
\begin{eqnarray}
\left[ 1- \xi \phi ^{2} \right] G_{\mu \nu }
+g_{\mu \nu } \Lambda
 = &
\left( 1 - 2\xi \right) \nabla _{\mu } \phi \nabla _{\nu }
\phi
+\left(2 \xi -\frac {1}{2} \right) g_{\mu \nu } \left(
\nabla \phi \right)^{2} ;
\nonumber \\ &
-2\xi \phi \nabla _{\mu } \nabla _{\nu } \phi
+2\xi g_{\mu \nu } \phi \nabla ^{\rho }\nabla _{\rho } \phi ;
\nonumber \\
\label{eq:Einstein}
\end{eqnarray}
and the scalar field equation:
\begin{equation}
 \nabla _{\mu } \nabla ^{\mu } \phi  =  \xi  R\phi .
\label{eq:phi}
\end{equation}
It is useful to take the trace of the Einstein equations
(\ref{eq:Einstein}) to give the Ricci scalar:
\begin{equation}
R = \frac {\left( 1 - 6 \xi \right) \left( \nabla \phi \right) ^{2}
 + 4\Lambda }{1 -
\xi \left( 1 - 6 \xi \right) \phi ^{2}} ,
\label{eq:Ricciscalar}
\end{equation}
where we have made use of the scalar field equation (\ref{eq:phi}) to
substitute for $\nabla _{\mu } \nabla ^{\mu } \phi $ arising in the
expression for $R$.
This equation enables one to eliminate higher-order derivatives of the
metric from the scalar field equation (\ref{eq:phi}).

We consider a static, spherically symmetric black hole geometry with line
element
\begin{eqnarray}
ds^{2} & = & -\left( 1- \frac {2m(r)}{r} - \frac {\Lambda r^{2}}{3}
\right) \exp (2\delta (r)) \, dt^{2}
\nonumber \\ & &
+ \left( 1 -\frac {2m(r)}{r} -\frac {\Lambda r^{2}}{3} \right) ^{-1}
dr^{2}
+ r^{2} \, d\theta ^{2}
+ r^{2} \sin ^{2} \theta \, d \varphi ^{2},
\label{eq:metric}
\end{eqnarray}
and we assume that the scalar field $\phi $ depends only on the
radial co-ordinate $r$.
For later convenience, we define the quantity $N(r)$ as
\begin{displaymath}
N(r)= 1 - \frac {2m(r)}{r} - \frac {\Lambda r^{2}}{3}.
\end{displaymath}
The Einstein equations for
this system are (a prime ${}'$ denotes $d/dr$):
\numparts
\begin{eqnarray}
\frac {2}{r^{2}} \left( 1- \xi \phi ^{2} \right) m' &  = &
\xi \Lambda \phi ^{2} + \left( \frac {1}{2} - 2 \xi \right) N \phi '^{2}
-\xi \phi \phi ' N' - 2 \xi N \phi \phi ''
\nonumber \\ & &
-\frac {4N}{r} \xi \phi \phi ' ;
\label{eq:Einsym1}
\\
\frac {2}{r} \left( 1 -\xi \phi ^{2} \right) \delta ' & = &
\left( 1- 2\xi \right) \phi '^{2} - 2 \xi \phi \phi ''
+ 2 \xi \phi \phi ' \delta '  ;
\label{eq:Einsym2}
\end{eqnarray}
\endnumparts
and the scalar field equation takes the form
\begin{equation}
N \phi '' + \left( N \delta ' + N' + \frac {2N}{r} \right)
\phi ' - \xi R \phi  =0.
\label{eq:phisym}
\end{equation}
It is now possible, using (\ref{eq:Ricciscalar}), to eliminate the
Ricci scalar curvature from the scalar field equation
(\ref{eq:phisym}) and then also eliminate $\phi ''$ from the
right-hand-side of the Einstein equations
(\ref{eq:Einsym1},\ref{eq:Einsym2}) above.
Writing the equations in this form would make them more amenable to
numerical integration.

We are interested in black hole solutions possessing a regular,
non-extremal event horizon at $r=r_{h}$, close to which the field
variables have the form
\begin{eqnarray*}
N(r) & = & N'(r_{h}) \left( r - r_{h} \right) + O(r-r_{h})^{2} ;
\nonumber \\
\delta (r) & = & \delta (r_{h}) + O(r-r_{h});
\nonumber \\
\phi (r) & = & \phi (r_{h}) + O(r-r_{h}).
\end{eqnarray*}
There may also be a cosmological event horizon at $r=r_{c}> r_{h}$
(depending upon the sign of the cosmological constant),
with similar expansions of the fields nearby.

At infinity, we assume that the geometry approaches asymptotically (anti-)de Sitter
space, and that the scalar field $\phi $ takes the form
\begin{equation}
\phi = \phi _{\infty } + O(r^{-k}) ,
\label{eq:phiinfinity}
\end{equation}
where $\phi _{\infty }$ is a constant,
for some $k>0$, and $k$ is not necessarily an integer.
Since we are working only with a vanishing potential $V$, we find, as in
\cite{ewc}, that consistency with the Einstein equations (\ref{eq:Einsym1},\ref{eq:Einsym2})
and the expression (\ref{eq:Ricciscalar}) for the Ricci scalar curvature requires
$\phi _{\infty }=0$ and then the scalar field equation (\ref{eq:phisym}) gives the
following equation for $k$:
\begin{displaymath}
k^{2} - 3k + 12 \xi =0,
\end{displaymath}
which has solutions
\begin{equation}
k = \frac {3}{2} \left( 1 \pm {\sqrt {1- \frac {16\xi }{3} }} \right) .
\label{eq:k}
\end{equation}
It is clear that $k$ has a positive real part for all $\xi >0$,
so that the scalar field $\phi $ converges to zero at infinity.
If $\xi > 3/16$, then $k$ is no longer real but has a non-zero imaginary part.
In this case, it is expected that $\phi $ oscillates about zero with decreasing amplitude
as $r \rightarrow \infty $, as was observed for a minimally coupled scalar field in
anti-de Sitter space with a non-zero self-interaction potential \cite{toriiAdS}.
If $\xi <0$, then one root of $k$ is negative, which means that the scalar field
$\phi $ diverges at infinity, although the other root for $k$ is positive.
We rule out the case in which $\phi $ diverges at infinity since, in that situation,
the geometry would no longer approach asymptotically (anti-)de Sitter space
in a manner compatible with the constraint (\ref{eq:Ricciscalar}) on the Ricci
scalar curvature.
Notice that the form of $k$ does not depend on the cosmological constant here (unlike
\cite{ewc}) because we have set the potential to zero.

Substituting the asymptotic behaviour of $\phi $ (\ref{eq:phiinfinity})
into the Einstein equation
(\ref{eq:Einsym2}), we find that $\delta ' \sim \Or (r^{-2k-1})$ for all $k$, so
$\delta \rightarrow \delta _{\infty } +  \Or (r^{-2k})$, for some constant
$\delta _{\infty }$, which we may take to be zero.
Using the other Einstein equation (\ref{eq:Einsym1}), we find that
$m'\sim \Or (r^{-2k+2})$.
When $\xi < 3/16$, the constant $k$ is real and,
taking the positive root in (\ref{eq:k}), $k>3/2$, so in this case,
$m \rightarrow M + \Or (r^{-2k+3})$ as $r\rightarrow \infty $, where $M$
is a constant,
and the metric function $m(r)$ converges to $M$ at infinity.
However, when $\xi \ge 3/16$, the constant $k$ is complex with real part equal to $3/2$,
which means that $m \sim \Or (\ln r) $ as $r \rightarrow \infty $.
A similar situation arises in \cite{logbranch}, for a minimally coupled scalar
field with a non-zero self-interaction potential.
The consequences of the logarithmic branch for the asymptotic structure of the
geometry are explored in \cite{logbranch}, and we shall not consider them further here.

\section{Simple No-hair Theorems}
\label{sec:nohair}

In this section we take the simple approach of Bekenstein \cite{bek72a}
to prove some elementary no hair theorems.
Similar techniques were used in \cite{ewc} to prove the non-existence of
non-trivial hairy black hole solutions when the cosmological constant
is positive and the scalar field conformally coupled.
It should be emphasized that this approach only rules out regular scalar field
hair (i.e. scalar fields such that the function $\phi $ is finite everywhere, including
at any event or cosmological horizon).
So, for example, the BBMB black hole \cite{bek74}-\cite{bbm} evades our proof
as the scalar field in that case diverges at the event horizon.

We start with the scalar field equation (\ref{eq:phisym}), multiply both
sides by $\phi r^{2} e^{\delta }$ and integrate from the event horizon
$r=r_{h}$ to $r=x$, where $x=r_{c}$, the radius of the cosmological horizon, if
$\Lambda >0$ and $x=\infty $ if $\Lambda <0$.
This gives the equation:
\begin{eqnarray}
0 & = &
\int _{r_{h}}^{x} dr
\left\{
r^{2} e^{\delta } \xi R \phi ^{2}
-
\phi
\left( Nr^{2} e^{\delta }
\phi ' \right) ' \right\}
\nonumber \\
& = &
-\left[ N r^{2} e^{\delta }
 \phi \phi '
\right] _{r_{h}}^{x} +
\int _{r_{h}}^{x} dr
\,
r^{2} e^{\delta } \left[ N \phi ^{'2}
+ \xi R \phi ^{2} \right] ,
\label{eq:parts}
\end{eqnarray}
where we have integrated by parts in the second line.
It is clear that the boundary term in (\ref{eq:parts}) vanishes at
a regular event or cosmological horizon where both $\phi $ and its derivative
are finite.
If $\Lambda <0$, the boundary term at infinity also vanishes if $\xi <0$ because
of the requirement imposed in section \ref{sec:model} that $\phi $ must tend to zero
at infinity, so we take the positive root for $k$ in (\ref{eq:k}).
This is sufficient for our requirements in section \ref{sec:case3} below, but we note
that the boundary term at infinity
also vanishes if $0<\xi < 3/16$, so that $k>3/2$ (\ref{eq:k})
but does not vanish if $\xi \ge 3/16$, in which case the real part of $k$ is equal to $3/2$.

Setting the boundary term equal to zero,
we now substitute, in equation (\ref{eq:parts}),
for the Ricci scalar $R$ from (\ref{eq:Ricciscalar}), to give
\begin{displaymath}
0 = \int _{r_{h}}^{x} r^{2} e^{\delta }
\left[
\frac {{\cal {F}}}{{\cal {G}}}
\right] ,
\end{displaymath}
where
\begin{eqnarray}
{\cal {F}} & = &
N \phi ^{'2} + 4 \Lambda \xi \phi ^{2} ;
\nonumber
\\
{\cal {G}} & = &
1 - \xi \left( 1- 6\xi \right) \phi ^{2}.
\label{eq:FandG}
\end{eqnarray}
It is then straightforward to prove no-hair theorems using the properties of
${\cal {F}}$ and ${\cal {G}}$, in the cases $\Lambda >0$, $\xi >0$ and $\Lambda <0$, $\xi <0$.
We now consider these cases in turn.

\subsection{$\Lambda >0$ and $\xi >1/6$}
\label{sec:case1}

In this case there is a cosmological horizon at $r=r_{c}$ and
we have from (\ref{eq:FandG}) that both ${\cal {F}}$ and ${\cal {G}}$
are positive and finite everywhere between $r_{h}$ and $r_{c}$.
Therefore, it must be the case that ${\cal {F}}\equiv 0$ for all $r\in [r_{h},r_{c}]$.
This is only possible if $\phi '$ and $\phi $ are identically zero,
so the black hole has no hair and the geometry is the trivial
Schwarzschild-de Sitter black hole.

\subsection{$\Lambda <0$ and $\xi <0$}
\label{sec:case2}

The fact that the black hole can have no non-trivial scalar field hair in this case
follows by exactly the same argument as in subsection \ref{sec:case1},
as here it is also the case that both ${\cal {F}}$ and ${\cal {G}}$ are positive
everywhere outside the event horizon, and we have shown at the beginning of section
\ref{sec:nohair} that the boundary term at infinity vanishes in this case.

\subsection{$\Lambda >0$ and $0<\xi <1/6$}
\label{sec:case3}

This case is slightly more complicated because here it might be possible for
${\cal {G}}$ to vanish somewhere between the event and cosmological horizon.
Suppose that, at some point $r=r_{0}$, the function ${\cal {G}}$ has a zero.
This point will be a curvature singularity unless ${\cal {F}}$ has a zero there as well.
Now ${\cal {F}}$ is the sum of two positive terms for these values of $\Lambda $ and $\xi $,
so, in order for ${\cal {F}}$ to vanish, each of these positive terms must be zero at this
particular value of $r$.
Therefore it must be the case that $\phi =0$ at $r=r_{0}$.
However, we then have a contradiction as $\phi =0$ implies that ${\cal {G}}=1$
at $r=r_{0}$, whereas we started with the assumption that ${\cal {G}}=0$
at $r=r_{0}$.
Hence ${\cal {G}}$ can have no zeros between the event and cosmological horizon, and
is of one sign.

It does not matter which sign ${\cal {G}}$ has, as we see that ${\cal {F}}$ is positive
everywhere, and so the only possibility we are left with is that ${\cal {F}}$ vanishes
identically between the event and cosmological horizon.
This is then the same as the two previous cases, and the black hole has no non-trivial
scalar field hair.

\subsection{Other Cases}
\label{sec:others}

We can add our results in this section to the table introduced in section \ref{sec:intro}
(new results are highlighted in bold):
\begin{center}
\begin{tabular}{|c||c|c|c|c|c|}
\hline
V=0 & $\xi <0 $ & $\xi =0$ & $0< \xi < 1/6$ & $\xi =1/6$ & $\xi > 1/6 $
\\
\hline
\hline
\hline
$\Lambda >0$ &  & No hair & {\bf {No hair}} & No hair &
{\bf {No hair}}
\\
\hline
$\Lambda =0$ & No hair & No hair & No hair & Unstable hair & No hair
\\
\hline
$\Lambda <0$ & {\bf {No hair}} & No hair &  & Stable hair &
\\
\hline
\end{tabular}
\end{center}
Unfortunately the simple arguments employed in this section cannot be extended to the
remaining cases (the blanks in the table) as it is no longer straightforward to show
that ${\cal {F}}$ and ${\cal {G}}$ are of one sign outside the black hole event horizon.
Furthermore, in this section the fact that we have been considering only vanishing potential
has been crucial.
For non-zero potential, this method is not useful (as was the case for $\Lambda =0$ and
$\xi =0$ \cite{bek95}), although it may be possible to extend it to a limited number
of potentials $V$ (for example, this was done for a quadratic potential in the
$\Lambda >0$, $\xi = 1/6$ case in \cite{ewc}).

\section{Conformal Transformation}
\label{sec:conformal}

In our previous work on conformally coupled scalar fields \cite{ewc},
we found it useful to employ a conformal transformation \cite{bek74,maeda},
which mapped the system onto the much simpler system involving just
a minimally coupled scalar field.
We shall apply the same techniques in this section.

\subsection{Definition of the Conformal Transformation}
\label{sec:confdef}

The system described by the action (\ref{eq:action})
above can be transformed
under a conformal transformation as follows \cite{bek74,maeda}:
\begin{equation}
{\bar {g}}_{\mu \nu } =\Omega g_{\mu \nu } ,
\label{eq:metricT}
\end{equation}
where
\begin{displaymath}
\Omega = 1-\xi \phi ^{2} .
\end{displaymath}
This transformation is valid only for those solutions for which
$\Omega $ does not vanish, which is automatically the
case if $\xi <0$, but will be a non-trivial assumption for $\xi >0$.
Under this transformation the action becomes that of a minimally
coupled scalar field:
\begin{equation}
S=\int d^{4}x \, {\sqrt {-{\bar {g}}}} \left[
\frac{1}{2} \left( {\bar {R}}-2\Lambda  \right)
-\frac {1}{2} \left( {\bar {\nabla }} \Phi \right) ^{2}
-U(\Phi ) \right] ,
\label{eq:actionT}
\end{equation}
where a bar denotes quantities calculated using the transformed
metric ${\bar {g}}_{\mu \nu }$, and we have defined a new scalar
field $\Phi $ as \cite{maeda}:
\begin{equation}
\Phi =\int d\phi \left[
\frac {1-\xi (1-6\xi )\phi ^{2}}{
(1-\xi \phi ^{2})^{2}} \right] ^{\frac {1}{2}} .
\label{eq:Phidef}
\end{equation}
For all values of $\xi $, we will choose the constant of
integration so that $\Phi =0$ when $\phi =0$.
For values of $\xi $ not equal to $1/6$ (or zero), the field $\Phi $
(\ref{eq:Phidef}) can be written in terms of inverse
sinh and tanh functions of $\phi $, but cannot be readily inverted
in simple closed form.
The transformed potential $U(\Phi )$ takes the implicit form \cite{maeda}
\begin{equation}
U(\Phi ) = \frac {\Lambda \xi \phi ^{2} \left(
2- \xi \phi ^{2} \right) }{ \left( 1- \xi \phi ^{2} \right) ^{2} } .
\label{eq:Upotential}
\end{equation}
Note that the presence of the cosmological constant means that the potential
in the minimally coupled scalar field system is not zero, even though the potential
in the non-minimally coupled scalar field system is zero.
We also note that, although the cosmological constant introduces a length scale into
the theory, it is unchanged by this transformation.

We assume that the transformed metric ${\bar {g}}_{\mu \nu }$ (\ref{eq:metricT})
is spherically symmetric, and we take it to have the form:
\begin{displaymath}
d{\bar {s}}^{2} = -{\bar {N}}({\bar {r}})
e^{2{\bar {\delta }}({\bar {r}})} dt ^{2}
+ {\bar {N}}({\bar {r}}) ^{-1} d{\bar {r}}^{2}
+ {\bar {r}}^{2} \left( d\theta ^{2} +
\sin \theta ^{2} \, d\varphi ^{2} \right) ,
\end{displaymath}
where we have defined a transformed radial co-ordinate $r$ by
\begin{displaymath}
{\bar {r}} = \left( 1 - \xi \phi ^{2} \right) ^{\frac {1}{2}} r.
\end{displaymath}
We also define a new quantity ${\bar {m}}({\bar {r}})$ by
\begin{displaymath}
{\bar {N}}({\bar {r}}) = 1
- \frac {2{\bar {m}}({\bar {r}})}{{\bar {r}}}
-\frac {\Lambda {\bar {r}}^{2}}{3};
\end{displaymath}
in terms of which the field equations derived from the transformed
action (\ref{eq:actionT}) are:
\numparts
\begin{eqnarray}
\frac {d{\bar {m}}}{d{\bar {r}}} & = &
\frac {{\bar {r}}^{2}}{4}  {\bar {N}}
\left(\frac {d\Phi }{d{\bar {r}}}\right) ^{2}
+ \frac {{\bar {r}}^{2}}{2}  U(\Phi ) ;
\label{eq:minE1}
\\
\frac {d{\bar {\delta }}}{d{\bar {r}}} & = &
\frac {{\bar {r}}}{2} \left( \frac {d\Phi }{d{\bar {r}}} \right) ^{2} ;
\label{eq:minE2}
\\
0 & = &
{\bar {N}} \frac {d^{2}\Phi }{d{\bar {r}}^{2}}
+ \left( {\bar {N}} \frac {d{\bar {\delta }}}{d{\bar {r}}}
+ \frac {d{\bar {N}}}{d{\bar {r}}}
+ \frac {2{\bar {N}}}{{\bar {r}}} \right)
\frac {d\Phi }{d{\bar {r}}}
- \frac {dU}{d\Phi } .
\label{eq:mincscal}
\end{eqnarray}
\endnumparts
Further details of the conformal transformation and the relationship between
the original and the transformed metric can be found in \cite{ewc}.

We now use the conformal transformation
to prove a no-hair theorem in the case $\Lambda >0$ and $\xi <0$.

\subsection{Application of the Conformal Transformation: $\Lambda >0$ and $\xi <0$}
\label{sec:confapp}

\begin{figure}
\begin{center}
\includegraphics[angle=270,width=8cm]{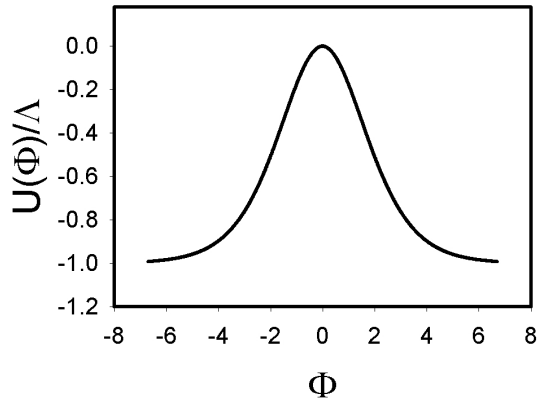}
\caption{{\label {fig:Upotential}}
The transformed potential $U/\Lambda $ given by (\ref{eq:Upotential}) in the
case $\Lambda >0$ and $\xi =-0.1$.
The form of the potential is very similar for other negative values of $\xi $.}
\end{center}
\end{figure}

We saw in section \ref{sec:nohair} that the simple techniques employed there to
prove no-hair theorems did not apply to this case.
We shall therefore consider a different approach, and employ the conformal
transformation defined in the previous subsection.
As $\xi <0$, the conformal transformation is well-defined provided the scalar
field $\phi $ remains regular.
In this case the potential of the minimally coupled scalar field $\Phi $ is
negative for all values of $\Phi $ and has a single stationary point, a maximum,
when $\Phi =0$ (see figure \ref{fig:Upotential}, where we plot the potential $U/\Lambda $
for the case $\xi =-0.1$).

Unlike the previous section, in this case we concentrate on the behaviour of the
scalar field outside the cosmological horizon in order to prove that only
the trivial solution is possible.
Recall that at the cosmological horizon, ${\bar {N}}=0$ and
$\frac {d {\bar {N}}}{d {\bar {r}}} <0$ because ${\bar {N}}<0$ outside the
cosmological horizon.
At the cosmological horizon, using (\ref{eq:mincscal}), we have
\begin{displaymath}
\frac {d{\bar {N}}}{d {\bar {r}}} \frac {d\Phi }{d{\bar {r}}} = \frac {dU}{d\Phi }.
\end{displaymath}
Suppose, firstly, that $\Phi >0$ at the cosmological horizon, then $\frac {dU}{d\Phi }<0$
and therefore $\frac {d\Phi }{d{\bar {r}}} >0$, so that $\Phi $ is positive
and increasing at the cosmological horizon.
Now, we know from the discussion in section \ref{sec:model} that the
non-minimally coupled scalar field $\phi $
must tend to zero at infinity, and, this is then also the case for the minimally coupled
scalar field $\Phi $.
Therefore $\Phi $ must have a maximum somewhere outside the cosmological horizon,
at which point $\frac {d\Phi }{d {\bar {r}}}=0$.
Substituting this into the field equation (\ref{eq:mincscal}) gives the equation,
\begin{displaymath}
{\bar {N}} \frac {d^{2} \Phi }{d {\bar {r}}^{2}} = \frac {dU}{d\Phi }.
\end{displaymath}
Bearing in mind that ${\bar {N}}<0$ because we are outside the cosmological horizon,
and that $\frac {dU}{d\Phi }<0$ because $\Phi >0$, we see that at this stationary point,
\begin{displaymath}
\frac {d^{2} \Phi }{d {\bar {r}}^{2}} >0 ,
\end{displaymath}
and so any stationary point of $\Phi $ outside the cosmological horizon can only
be a minimum when $\Phi >0$.
Therefore it is not possible for $\Phi $ to have a maximum and so the
boundary condition at infinity cannot be satisfied.
The argument carries over similarly if $\Phi $ is negative at the cosmological horizon,
only in this case $\Phi $ is decreasing at the cosmological horizon but is unable to
have a minimum outside the cosmological horizon and thus cannot satisfy the boundary
condition at infinity.

The only possibility is therefore that $\Phi =0$ at the cosmological horizon.
By repeatedly differentiating the scalar field equation (\ref{eq:mincscal}), it can be
shown that all derivatives of $\Phi $ then vanish at the cosmological horizon, and
we deduce that $\Phi \equiv 0$ everywhere.
Adding this result to our table, we now have:
\begin{center}
\begin{tabular}{|c||c|c|c|c|c|}
\hline
V=0 & $\xi <0 $ & $\xi =0$ & $0< \xi < 1/6$ & $\xi =1/6$ & $\xi > 1/6 $
\\
\hline
\hline
\hline
$\Lambda >0$ & {\bf {No hair}} & No hair & {\bf {No hair}} & No hair &
{\bf {No hair}}
\\
\hline
$\Lambda =0$ & No hair & No hair & No hair & Unstable hair & No hair
\\
\hline
$\Lambda <0$ & {\bf {No hair}} & No hair &  & Stable hair &
\\
\hline
\end{tabular}
\end{center}

It should be emphasized that our no-hair theorem in this case applies only for
scalar field hair which is a function of the co-ordinate $r$ only, with the
metric of the form (\ref{eq:metric}).
This is a strong assumption, particularly in the region exterior to the
cosmological horizon, where $r$ is a time-like co-ordinate.
We are unable to rule out black hole solutions for which the scalar field is not
just a function of $r$ outside the cosmological horizon, or for which either $\phi $ or
one of its derivatives ceases to be regular on the cosmological horizon.
However, this does not render our result without significance;
for example,
it rules out black hole solutions similar to those presented in \cite{martinez}, where
there is a scalar field depending only
on the co-ordinate $r$, including in the region outside the
cosmological horizon.
Similarly, in the
Einstein-Yang-Mills system, non-trivial solutions do exist when the metric outside
the cosmological horizon has the form (5), with the matter fields depending
only on the co-ordinate $r$ \cite{volkov}.

\section{Existence and Stability of Hairy Black Holes when $\Lambda <0$ and $\xi >0$}
\label{sec:existence}

The only region of parameter space where we have not been able to prove a no-hair
theorem is when $\Lambda <0$ and $\xi >0$.
Given that there are known to be stable, non-trivial, solutions when $\xi =0$ or $1/6$,
it perhaps comes as no surprise to find hairy black hole solutions in this case.

\subsection{Existence of Hairy Black Hole Solutions}
\label{sec:solns}

To find the solutions, we integrated the minimally coupled field equations
(\ref{eq:minE1}--\ref{eq:mincscal})
numerically, and then transformed these solutions back to the non-minimally coupled system.
This is the same strategy as employed for the conformally coupled scalar field system in \cite{ewc},
but means that we only find those solutions for which the conformal transformation (\ref{eq:metricT}) is
valid.
The solutions display very similar properties to those found in the conformally coupled
case \cite{ewc}, with the scalar field $\phi $ monotonically decaying to zero from its value
on the event horizon when $\xi <3/16$.
When $\xi > 3/16$,
the scalar field starts to oscillate about zero with decreasing amplitude, as predicted
in section \ref{sec:model}.

\begin{figure}
\begin{center}
\includegraphics[angle=270,width=10cm]{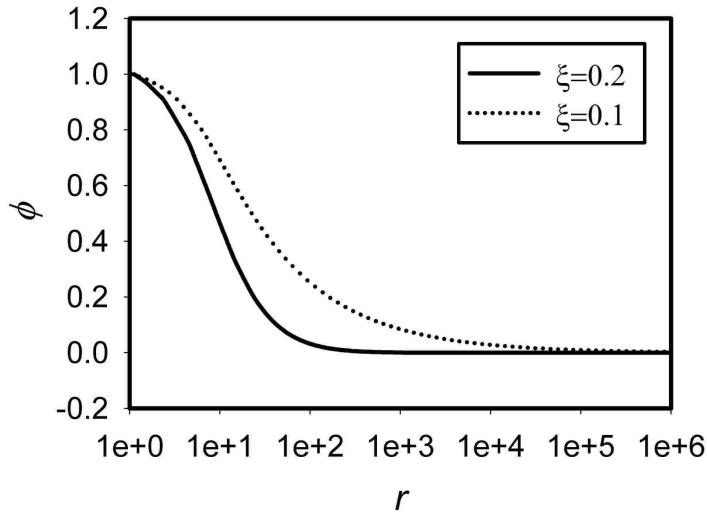}
\caption{{\label{fig:phi}}
Examples of typical hairy black hole solutions with a non-minimally coupled
scalar field, when $\xi =0.1$ (dotted) and $\xi =0.2$ (solid).
For these solutions, the event horizon radius is taken to be
$r_{h}=1.0541$ for the $\xi =0.1$ solution and
$r_{h}=1.118$ for the $\xi =0.2$ solution,
the cosmological constant $\Lambda = -0.1$ and the value of the scalar field at the
event horizon is $\phi (r_{h})=1$.
Solutions for other values of the parameters $\Lambda $, $r_{h}$, $\xi $ and $\phi (r_{h})$
behave similarly.}
\end{center}
\end{figure}

\begin{figure}
\begin{center}
\includegraphics[angle=270,width=12cm]{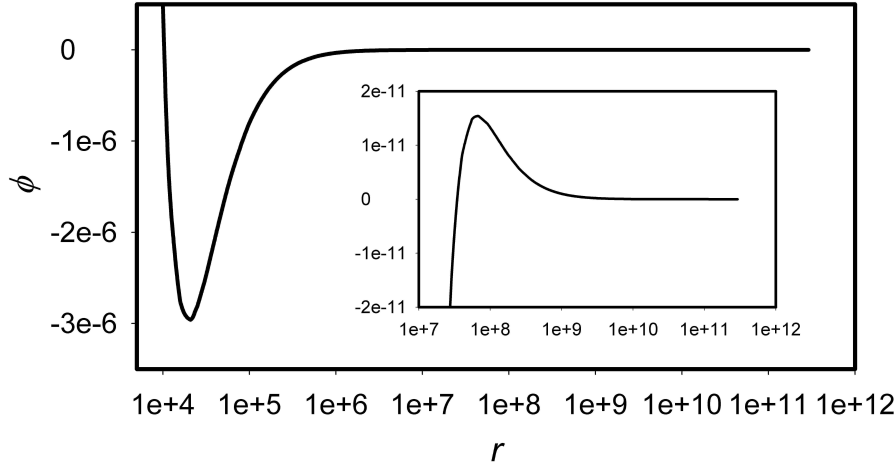}
\caption{{\label{fig:phiosc}}
Example of a typical hairy black hole solution with a non-minimally coupled
scalar field, when $\xi =0.2>3/16$.
The values of the other parameters are as in figure \ref{fig:phi}.
The first oscillation in $\phi $ about zero can be seen in the main graph, while the inset shows
the second oscillation.}
\end{center}
\end{figure}

As an example, two typical solutions are shown in figure \ref{fig:phi}, for
the values $\xi =0.1 $ (dotted) and $\xi =0.2 $ (solid), where the cosmological constant
$\Lambda =-0.1$, the radius of the event horizon $r_{h}=1.0541$ for the $\xi =0.1$ solution
and $r_{h}=1.118$ for the $\xi =0.2$ solution, and the value of the scalar
field on the event horizon is $\phi (r_{h})=1$.
We have investigated a large number of numerical solutions by varying the parameters
$\Lambda $, $r_{h}$, $\xi $, and $\phi (r_{h})$ and found the same qualitative behaviour
in each case.
From figure \ref{fig:phi}, the monotonically decreasing behaviour of $\phi $ as $r\rightarrow \infty $
can be seen for $\xi <3/16$.
For $\xi >3/16$, the oscillatory behaviour of $\phi $ as $r\rightarrow \infty $ can be seen in figure
\ref{fig:phiosc}.

\subsection{Stability of the Hairy Black Hole Solutions}
\label{sec:stab}

We now investigate the stability of these hairy black holes, using the same
approach as in \cite{ewc}.
We consider linear, spherically symmetric perturbations of the metric and scalar
field.
The algebra is simplest if we make use of the conformal transformation, but
either way it is straightforward to eliminate the perturbations of the metric functions
from the linearized Einstein equations and obtain a single perturbation equation
for the quantity $\psi $, which is related to the perturbation of the scalar field,
$\delta \phi $, by
\begin{displaymath}
\psi =
r \left( 1- \xi \phi ^{2} \right) ^{-\frac {1}{2}}
\left[ 1- \xi (1-6\xi ) \phi ^{2} \right] ^{\frac {1}{2}}
\delta \phi  .
\end{displaymath}
This single perturbation equation has the standard Schr\"odinger form,
for perturbations which are periodic in time (i.e. $\psi (t,r)=e^{i\sigma t}\psi (r)$):
\begin{equation}
 \sigma ^{2} \psi  =
- \frac {\partial ^{2} \psi }{\partial r^{2}_{*}} +
{\cal {U}} \psi ,
\label{eq:pert}
\end{equation}
where $r_{*}$ is the usual ``tortoise'' co-ordinate given in terms of the radial co-ordinate
$r$ by
\begin{displaymath}
\frac {dr_{*}}{dr} = \frac {e^{-\delta }}{N};
\end{displaymath}
and, because we are working in asymptotically anti-de Sitter space, the
tortoise co-ordinate $r_{*}$, by a choice of
the constant of integration, lies in the interval $( - \infty ,0]$.
The perturbation potential ${\cal {U}}$ is given by:
\begin{eqnarray}
{\cal {U}} & = &
\frac { N e^{2\delta } } {r^{2}} \left[
1 - N {\cal {A}}^{-2} {\cal {B}} ^{2}
- \Lambda r^{2} {\cal {A}}
- \Lambda \xi r^{2} \phi ^{2} {\cal {A}}^{-1} \left( 2- \xi \phi ^{2} \right)
\right.
\nonumber \\ & &
\left.
+ 8 \Lambda \xi r^{3} \phi \phi ' {\cal {A}}^{-1} {\cal {B}} ^{-1}
+ 4 \Lambda \xi r^{2} {\cal {C}}^{-2}
+ 16 \Lambda \xi ^{2} r^{2} \phi ^{2} {\cal {A}}^{-1} {\cal {C}}^{-1}
\right.
\nonumber \\ & &
\left.
+ \Lambda r^{4} \phi ^{'2} {\cal {A}}^{-1} {\cal {B}}^{-2} {\cal {C}}
- r^{2} \phi ^{'2} {\cal {B}}^{-2} {\cal {C}}
\right] ;
\label{eq:pertpot}
\end{eqnarray}
where
\begin{eqnarray*}
{\cal {A}} & = &
1- \xi \phi ^{2} ;
\\
{\cal {B}} & = &
1- \xi \phi ^{2} - \xi r \phi \phi ' ;
\\
{\cal {C}} & = &
1 - \xi (1-6\xi ) \phi ^{2}.
\end{eqnarray*}
The perturbation potential ${\cal {U}}$ (\ref{eq:pertpot})
vanishes at the event horizon $r=r_{h}$, and
at infinity its leading order behaviour is
\begin{displaymath}
{\cal {U}} \sim \frac {2\Lambda ^{2}r^{2}}{9} \left( 1- 6\xi \right)  .
\end{displaymath}
This means that the potential diverges to positive infinity
like $+r^{2}$ as $r\rightarrow \infty $
for $\xi < 1/6$, and diverges to negative infinity like $-r^{2}$ if
$\xi > 1/6$.
In the case that $\xi =1/6$, the potential remains bounded at infinity, as
found in \cite{ewc}.

\begin{figure}
\begin{center}
\includegraphics[angle=270,width=10cm]{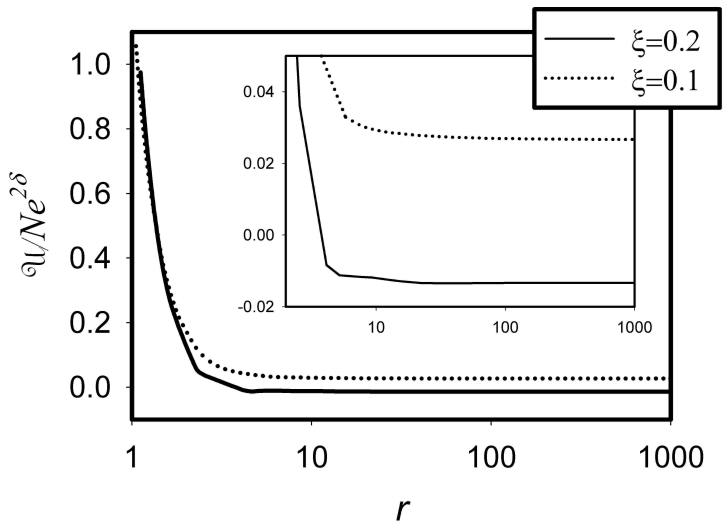}
\caption{{\label{fig:pertpot}}
Examples of the perturbation
potential ${\cal {U}}/Ne^{2\delta }$ for the
black hole solutions shown in figure \ref{fig:phi},
plotted as functions of the radial co-ordinate $r$.
The inset shows that the potential for $\xi =0.2>1/6$ (solid) is negative
for sufficiently large $r$, whereas the potential for $\xi =0.1<1/6$ (dotted)
is positive everywhere outside the event horizon.}
\end{center}
\end{figure}

As in \cite{ewc}, it is necessary to examine the potential numerically, for example,
in figure \ref{fig:pertpot} we plot, as a function of the radial co-ordinate $r$,
 the perturbation potential ${\cal {U}}$ (\ref{eq:pertpot})
divided by $Ne^{2\delta }$ for the black hole solutions shown in figure \ref{fig:phi}.
It can be seen that the potential for the solution with $\xi <1/6$ is positive everywhere
outside the event horizon, whereas as that for the solution with $\xi >1/6$ becomes
negative for sufficiently large $r$.
Similar behaviour was observed for all the other solutions we examined.

For those solutions with $\xi <1/6$ for which the perturbation potential ${\cal {U}}$ is
positive everywhere outside the event horizon, we can conclude immediately that
the perturbation equation (\ref{eq:pert}) has no bound state solutions and so
the black holes are linearly stable.

\begin{figure}
\begin{center}
\includegraphics[angle=270,width=8cm]{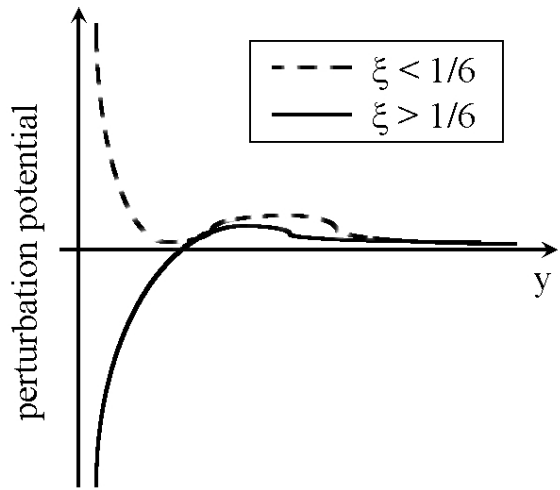}
\caption{{\label{fig:ypotsketch}}
Sketch of the perturbation potential ${\cal {U}}$ as
a function of $y=-r_{*}$ for the black hole solutions shown in figure \ref{fig:phi}.}
\end{center}
\end{figure}

The situation for the black holes with $\xi > 1/6$ is more complex, with the potential
diverging to $-\infty $ as $r_{*} \rightarrow 0$, and further analysis is needed.
If we define a new variable $y$ by $y=-r_{*}$, then $y\in [0, \infty )$, and the form
of the potential is sketched in figure \ref{fig:ypotsketch}.
We will examine the form of the zero mode solutions to
(\ref{eq:pert}).
The zero modes are the time-independent solutions of the perturbation equation
(\ref{eq:pert}).
There is a standard result (discussed, for example, in \cite{bargmann}),
that the number of bound states of the equation
\begin{displaymath}
\sigma ^{2} \psi = - \frac {\partial ^{2} \psi }{\partial y^{2}} + {\cal {U}}\psi
\end{displaymath}
(which is the same as the perturbation equation (\ref{eq:pert}), but with $y$ as the
independent variable instead of $r_{*}$)
is equal to the number of zeros of the zero mode $f(y)$ such that $f(0)=0$.

However, numerically it is easier to find the zero mode $g(r)$ of the
equivalent perturbation equation
\begin{equation}
-N^{2} e^{2\delta } g'' - N e^{\delta } \left( N e^{\delta } \right) ' g'
+{\cal {U}} g =0 ,
\label{eq:numpert}
\end{equation}
with suitable initial conditions on $g$ at the event horizon of the black hole
(that is, at $y\rightarrow \infty $).
By examining the form of the solutions to (\ref{eq:numpert}) as $r\rightarrow \infty $
(which corresponds to $y\rightarrow 0$),
it can be shown that
as $r\rightarrow \infty $, the function $g$ behaves like $r^{-\ell }$, where
\begin{equation}
\ell = \frac {1}{2} \left[ 1 \pm {\sqrt {9-48\xi }} \right] .
\label{eq:ell}
\end{equation}
The real part of $\ell $ is positive whenever $\xi >1/6$,
which is the case in which we are most interested here, so that $g(r)$ tends to zero
as $r \rightarrow \infty $.
Therefore this $g(r)$ constructed numerically turns out to be a valid zero mode $f(y)$.
Therefore, the number of bound state solutions of (\ref{eq:pert}) is equal to the
number of zeros of our numerical function $g$.

\begin{figure}
\begin{center}
\includegraphics[angle=270,width=10cm]{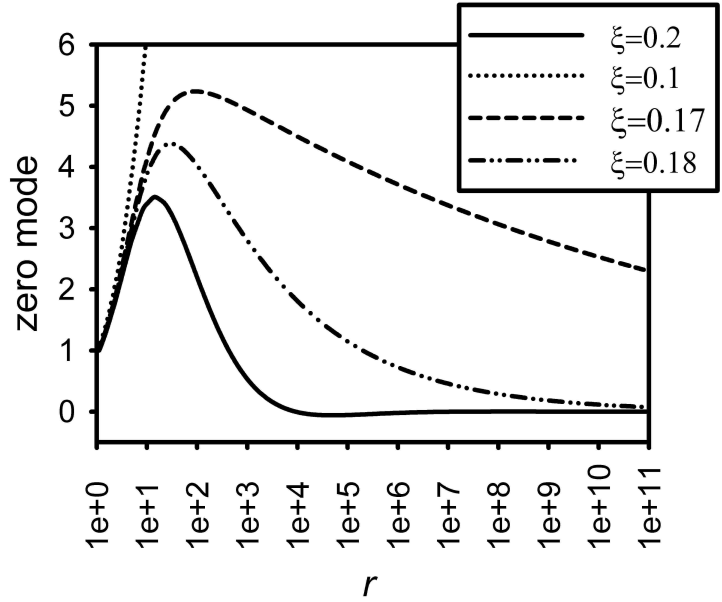}
\caption{{\label {fig:zeromodes}}
The zero mode solutions of the perturbation equation (\ref{eq:numpert})
for the equilibrium black hole solutions plotted in
figure \ref{fig:phi}, and the corresponding solutions when $\xi =0.17$ and $0.18$.
The cosmological constant is $\Lambda =-0.1$.}
\end{center}
\end{figure}

\begin{figure}
\begin{center}
\includegraphics[angle=270,width=12cm]{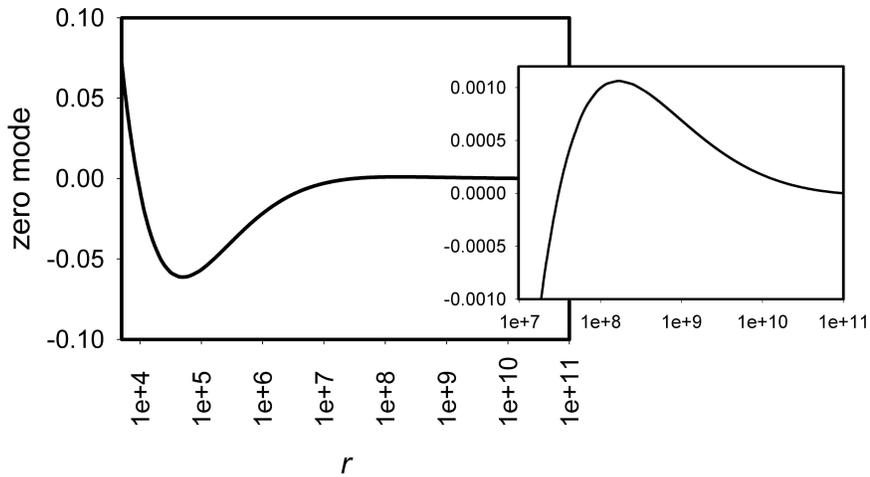}
\caption{{\label {fig:zeroosc}}
Detailed structure of the zero mode solution of the perturbation equation (\ref{eq:numpert})
for $\xi =0.2$, $\phi (r_{h}) =1$ and $\Lambda = -0.1$.
The main graph shows the first oscillation in the zero mode; the inset the second.}
\end{center}
\end{figure}

In figure \ref{fig:zeromodes} we plot the corresponding zero mode function $g$ for the
black hole solutions shown in figure \ref{fig:phi}, and also for similar black hole solutions
(with $\phi (r_{h})=1$) when $\xi =0.17$ and $0.18$.
We find that, for $\xi < 1/6$, the zero mode functions have no zeros, and
simply increase away from their value at the event horizon.
For $\xi > 1/6$, the situation is more complicated, depending on whether $\xi \le 3/16$ or
$\xi > 3/16$.
If $1/6<\xi \le 3/16$, then from (\ref{eq:ell}), the constant $\ell $ is real and
the zero mode function $g$ monotonically decreases to zero as $r\rightarrow \infty $, as
can be seen in figure \ref{fig:zeromodes}, although the decrease to zero is slow since
$\ell <1/2$.
However, for $\xi > 1/6$, the zero mode functions have at least one zero,
and may in fact oscillate many times with decreasing amplitude as
$y \rightarrow 0$ ($r \rightarrow \infty $).
This can be seen in figure \ref{fig:zeroosc}, where we zoom in on the behaviour
of the zero mode when $\xi =0.2$, for the black hole solution shown in figure \ref{fig:phi}.
This means that for $\xi >3/16$ there is at least one bound state solution of (\ref{eq:pert}) with
$\sigma ^{2}<0$ and so the black holes are unstable.

\section{Conclusions}
\label{sec:conc}

In this paper we have studied black hole solutions in four-dimensional
general relativity with a scalar field which is non-minimally coupled to the Ricci
scalar curvature but otherwise has no self-interaction.

Using a simple technique due to Bekenstein \cite{bek72a}, we were able to rule
out non-trivial scalar field hair when $\Lambda \xi >0$, at least for this
zero potential case.
We expect that this result may be true also for some non-zero potentials,
for example, for a quadratic potential as in the conformally coupled case \cite{ewc}.
However, we do not expect it to be true for all potentials.
For example, for the minimally coupled scalar field, although there is no hair when
the potential vanishes, non-trivial scalar field hair does exist for the double-well
Higgs potential.
It does seem reasonable to conjecture that, in analogy with the minimally coupled case,
any scalar field hair when the cosmological constant is positive will be unstable, but there
may well be stable scalar field hair for negative cosmological constant.

We then used a conformal transformation \cite{bek74,maeda}, valid for $\xi <0$,
to rule out scalar field hair when $\Lambda >0$ and $\xi <0$.
Again, the vanishing of the self-interaction potential seems to be crucial in this
proof and we suspect that there may well be hair for some non-zero potentials in this case
(although we would again conjecture that such hair must be unstable).

Finally, when $\Lambda <0$ and $\xi >0$, we found numerically hairy black hole solutions
of the field equations, at least some of which are stable when $0<\xi <3/16$, but all the
solutions studied were unstable when $\xi >3/16$.
We summarize our results in the completed table below, with
new results derived in this paper are highlighted in bold:
\begin{center}
\begin{tabular}{|c||c|c|c|c|c|}
\hline
V=0 & $\xi <0 $ & $\xi =0$ & $0< \xi < 1/6$ & $\xi =1/6$ & $\xi > 1/6 $
\\
\hline
\hline
\hline
$\Lambda >0$ & {\bf {No hair}} & No hair & {\bf {No hair}} & No hair &
{\bf {No hair}}
\\
\hline
$\Lambda =0$ & No hair & No hair & No hair & Unstable hair & No hair
\\
\hline
$\Lambda <0$ & {\bf {No hair}} & No hair & {\bf {Stable hair}} & Stable hair &
{\bf {Unstable hair}}
\\
 & & & & & for $\xi > 3/16$
 \\
\hline
\end{tabular}
\end{center}
The instability of the black holes when $\xi >3/16$ may be understood in terms of the
Breitenlohner-Freedman bound \cite{breit1,breit2}.
In the asymptotically anti-de Sitter region of space-time, the scalar field equation
(\ref{eq:phisym}) takes the form
\begin{displaymath}
\nabla _{\mu } \nabla ^{\mu } \phi = 4 \xi \Lambda \phi ,
\end{displaymath}
so that $4\Lambda \xi $ acts as a (negative) ``mass''-squared term in this limit.
The Breitenlohner-Freedman bound states that for ``masses''-squared less than
\begin{displaymath}
m_{*}^{2} = \frac {3\Lambda }{4} ,
\end{displaymath}
in four dimensions,
the corresponding scalar fields
produce a perturbative instability of anti-de Sitter space-time.
In our case the Breitenlohner-Freedman bound corresponds to $\xi = 3/16$, and
therefore for larger values of $\xi $ the Breitenlohner-Freedman instability is
reflected in the instability of our black hole solutions.
It is interesting to note that this is the same value of $\xi $ at which the
scalar field starts to oscillate about zero as $r\rightarrow \infty $, so the instability
is in accord with the observation in \cite{toriiAdS} for the minimally coupled scalar
field case, that the number of unstable modes of the black hole configuration was
equal to the number of zeros of the scalar field.

\ack We would like to thank Robert Mann and Eugen Radu for helpful
discussions.
This work was supported by UK PPARC, grant reference number PPA/G/S/2003/00082.
We thank Oriel College, Oxford, for hospitality while this
paper was completed.

\end{document}